# Multiparameter Monitoring and Fault Indication Using Inductive Power Transfer System

K.P.Shaji[1],      I.Alsheba[2],      Y.A.Syed Khadar[3],      S.Kannan[4]

*Abstract— The paper aims at demonstrating communication capabilities of IPT. For this data communication is performed between two modules using the concept of IPT. IPT was deemed to be the best solution to the system houses a multi parameter acquisition module such as temperature, speed, voltage, current and data transfer from the motor. The receiver side is another microcontroller coupled to an inductive coil that gets the data and displays in the LCD. A brief background to IPT (Inductive Power Transfer) technology and its applications is given and the design criteria for the paper are defined in detail. To be accurate, IPT data communication helps to reduce unnecessary wire connections and data is transmitted without any touch. Further the paper can be enhanced by looking for fault analysis inside the motor. This can be done by analyzing various parameters of the motor. A novel two-way IPT communication system was designed, which worked on the concept of pulsing the system on and off to send data serially. The paper involves transmission of data through inductive flux without any contact between the two modules. Further as no frequency tunings or any calibration is required between different modules a single system can be used with multiple clients. This reduces a lot of hazards such as interference with other modules and RF transmitters in the vicinity.*

## I. INTRODUCTION

IPT (Inductive Power Transfer) is a relatively new concept which is becoming widely accepted as an effective method for power transfer. It utilizes the same principles as the common transformer, except that an air gap is present to enable the secondary side to move with respect to the primary. Although air gaps reduce the coupling efficiency of transformers significantly, the IPT system uses power electronics to 'tune'

**Manuscript received May 20, 2013**.
**K.P.Shaji**, Electronics & Communication, Periyar University/ Muthayammal College of Arts & Science, Namakkal, India, Mobile No – 9842997714 , shajibindhu2002@gmail.com

**I.Alsheba**, Electronics & Electrical Engineering, Anna University / KSR College of Technology, Namakkal, India, Mobile No – 9994389942, alshefa@gmail.com

**Y.A.Syed Khadar**, Electronics & Communication, Periyar University/ Muthayammal College of Arts & Science, Namakkal, India, Mobile No – 9994092852, syedkhadar_ya@yahoo.co.in.

**S.Kannan**, Electronics & Communication, Periyar University/ Muthayammal College of Arts & Science, Namakkal, India, Mobile No – 9943734799, skannanec@yahoo.co.in.

the secondary side pick-up to improve this coupling. This results in efficiencies of more than 85%, which is impressive when compared with the efficiency of a typical power transformer of around 98%. The result is a form of power transfer that is contact less, reliable, and unaffected by dust or chemicals. IPT has found uses in a number of commercial applications because of these attributes. Among the many applications of IPT the most useful ones are in the field of implantable biomedical devices. Biomedical devices that are implanted within the body can be powered using IPT.

Another application of IPT is in powering road studs that contain LED's (Light Emitting Diodes). A long cable that is buried under the road powers the studs, enabling the road lanes to be seen clearly at night and during adverse weather conditions. The IPT system can also be used for communication between the power supply and the secondary coil or pick-up. Both AM and FM systems can be implemented in low data transfer rate applications.

### A. Transmitter circuit diagram and description

IPT transmission consists of the message signal generator, 555IC and power converter. In most cases the message signal generator is a controller that transmits binary values such as ones and zeros depending upon the data to be transmitted. These binary values if fed to the inductor directly will not produce pulse because of the low frequency and it will act more or less as a DC signal though it is a pulsed waveform. In order to produce flux a high frequency signal must be applied to the inductor. This frequency must be at the resonant frequency of the inductor.

To generate this frequency an astable multivibrator is used. An IC555 is used to generate higher frequency. But this frequency is generated only when a control input is given to the pin4 of the multivibrator. This signal comes from the microcontroller. So in a stream of data transmitted from the mc, the IC555 produces pulses only when the signal is high at all other times it does not produce any signal. This is a sort of amplitude shift keying wherein the amplitude of the output pulses is present when message signal is one and it becomes zero when the message signal is zero. SO when ever the message signal is one there is a high frequency signal produced fro the astable multivibrtaor. The IC555 cannot drive a high current Inductor coil. For this a high power transistor is used to drive the Inductor. When ever signal from IC555 is high the transistor is switched on and the inductor produces flux at all other cases there is no flux generated.

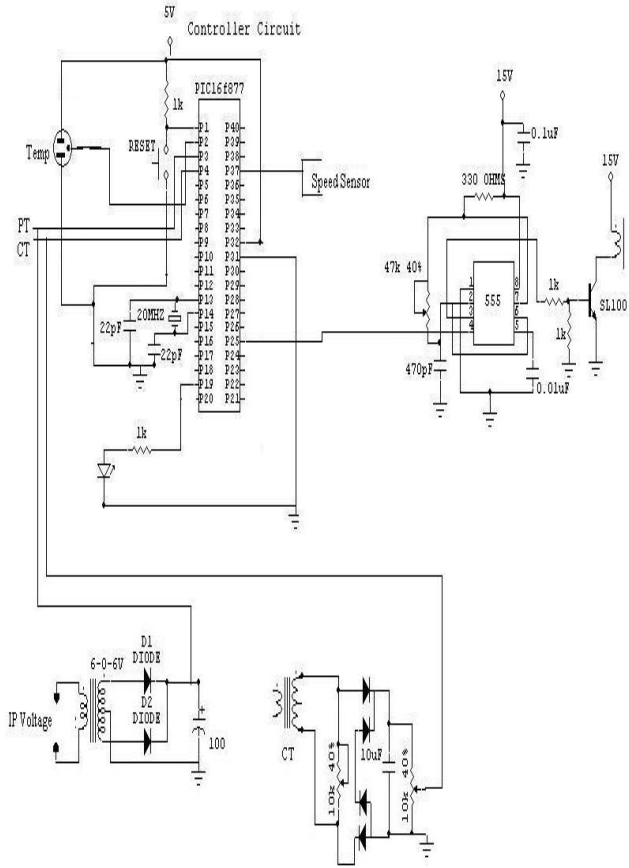

Fig 1. IPT Transmission Circuit

*B. Receiver circuit diagram and description*

IPT receiver consists of Inductive coil for reception, Noise filter and signal shaping circuit. Inductive coil directly receives the flux from the transmitter and converts it into electrical voltage. This voltage is a high frequency sine wave with a lot of noise in it. The first step is to reduce the noise present in it. For this a two level high frequency filter is used. This high frequency filter has a cut off frequency above the frequency of the IPT receiver. This is achieved by a parallel capacitor and series resistor in two stages. After this a clear signal is obtained which needs to be wave shaped to give it to the microcontroller. The wave shaping signal consists of a series capacitor and a parallel resistor which is a low pass filter that makes a sequence of pulsed sine wave into a constant high signal. Thus when there is a sine wave the signal is converted into a constant high and when there is no signal output is pure low. A level converter at the end makes this high and low into a constant 5v and 0v signal which is suitable for a microcontroller. The microcontroller receives the data signal for processing the pulsed signal through its serial port. Thus a two level filter circuit helps to reduce noise and also makes the sine wave signal into pulsed signal suitable for microcontroller.

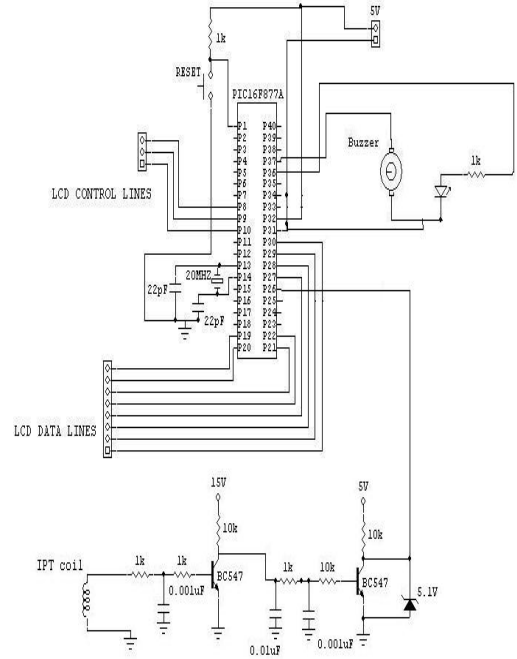

Fig 2. Receiver Circuit Diagram

**II. Speed Sensor: (Inductive proximity sensors)**

P+F inductive proximity sensors are the Eddy Current Killed Oscillator (ECKO) type. This type of sensor contains four basic elements as shown. The oscillator creates a radio frequency that is emitted from the coil away from the face of the sensor. If a metal plate enters this radiated field, eddy currents circulate within the metal. The oscillator requires energy to maintain the eddy currents in the metal plate. As the plate approaches the sensor, the eddy currents increase and cause a greater load on the oscillator. The oscillator stops when the load becomes too great. The trigger circuit senses when the oscillator stops, then changes the state of the switching device to control the load.

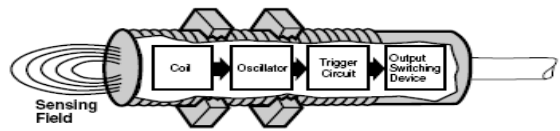

Fig 3. Speed Sensor Construction

The sensing field is in the front of the sensing device. The sensor is activated when a target enters the sensing field in an axial or lateral direction. The sensing range of a proximity sensor is determined by the size of the coil. Thus, the longer the sensing range, the larger the sensor. Factors that affect the sensing range include target size and composition, as well as ambient temperature.

*I. Sensor Specifications:*

- *Temperature range:*
  -14°F to +158°F Extended temperature range of -40°F to +212°F available on some models.
- *Repeatability:*
  ≤.01mm when tested at 75°F ±10% with a nominal supply voltage of VS ±5%.
- *Hysteresis:*
  0.03mm to 3mm for inductive sensors
  0.03mm to 10mm for capacitive sensors
- *Shock:* (IEC 68-2-6)
  Sine wave, acceleration 30 times gravitational constant, less than 11 milliseconds, 3 shocks in both directions (Forward and back), in all three planes, X, Y and Z.
- *Vibration:* (IEC 68-2-27)
  Frequency range 10-55Hz, amplitude 1mm, cycle time 5 minutes, 30 minutes in each plane X, Y and Z.

### III. Switching circuit

A switching circuit is a device which can turn ON or OFF current in an electrical circuit. It is the most important part of a switching circuit. This paper using electronic switch.Electronics switch is a device, which can turn ON or OFF current in an electrical circuit with the help of electronic devices e.g. transistor, tubes, SCR.Electronic switch have become very popular because of their high speed of operation and absence of sparking. The operation of transistor as a switch can be beautifully illustrated.

*M.        Switching transistor*

When the base input voltages is negative or zero the transistor is cut-off and no current flows in the load RC. As a result there is no voltage drop across the collector load and the output voltage is ideally Vcc. When the base input voltage is positive current IB flows in the base and amplified collector current IC flows through load RC.

$$Vout = Vcc – IcRc$$

The above discussion shows that a transistor can turn ON or OFF current in the load RC depending upon whether the input base voltage is positive or negative. Thus a transistor can act as a switch.

*A.  Advantages*

The following are the advantage of transistor switches over other types of switches.

1. It has no moving parts and hence there is little wear and tear.
2. Therefore it gives noiseless operation.
3. It has smaller size and weight.
4. It gives trouble free service because of solid state.
5. It is cheaper than other switches and requires little maintenance.
6. It has a very fast speed of operation.

### IV. LCD (liquid crystal display)

Liquid Crystal Displays (LCD's) have materials, which combine the properties of both Liquids and Crystals. Rather than having a melting point, they have a temperature range with in which the molecules are almost as mobile as they would be in a Liquid, but are grouped together in an ordered form similar to a crystal. An LCD consists of two glass panels, with the Liquid Crystal material sand witched in between them. The inner surfaces of the glass plates are coated with transparent electrodes, which define the character, symbols or patterns to be displayed. Polymeric layers are present in between the electrodes and the Liquid Crystal molecules to maintain a defined orientation angle.

On each polarizer's are pasted outside the two glass panels. This polarizer's would rotate the light rays passing through them to a definite angle, in a particular direction. When the LCD is in the off state, light rays are rotated by the two polarizer's Liquid Crystal, Such that the light rays come out of the LCD without any orientation, and hence the LCD appears transparent. When sufficient voltage is applied to the electrodes, the Liquid Crystal molecules would be aligned in a specific direction. The light rays passing through the LCD would be rotated by the polarizer's, which would result in activating / highlighting the desired characters.

The LCD's are lightweight with only a few millimeters thickness. Since the LCD's consume less power, they are compatible with low power electronic circuits, and can be powered for long durations. The LCD does don't generate light and so light is needed to read the display. By using backlighting, reading is possible in the dark. The LCD's have long life and a wide operating temperature range. Changing the display size or the layout size is relatively simple which makes the LCD's more customers friendly. The LCD's used exclusively in watches, calculators and measuring instruments are the simple seven-segment displays, having a limited amount of numeric data. The recent advances in technology have resulted in better legibility, more information displaying capability and a wider temperature range. These have resulted in the LCD's is being extensively used in telecommunications and entertainment electronics. The LCD's have even started replacing the Cathode Ray Tubes (CRTs) used for the display of text and graphics, and also in small TV applications.

*a. Back Lighting*

When sufficient lighting is not there, back lighting of the LCD is done for reading the characters/ patterns. The widely used back lighting is of the LED array type, where in the LEDs are connected in an array. The absence of any noise/interference, a typical life time of 100,000 hours on an average, the low DC drive voltage of 5V and the various colour options available makes the LED backlighting the widely used one. The Electro-optical characteristics of the LED array type are as given below: Since the sensitivity of eye is maximum at 550nm, the yellow-green back lighting colour is the most widely used one.

### b. PIC Micro Controller

The term PIC, or Peripheral Interface Controller, is the name given by Microchip Technologies to its single – chip microcontrollers. These devices have been phenomenally successful in the market for many reasons, the most significant ones are mentioned below. PIC micros have grown in steadily in popularity over the last decade, ever since their inception into the market in the early 1990s. PIC micros have grown to become the most widely used microcontrollers in the 8- bit microcontroller segment. The PIC16F877 is 40 pin IC. There are six ports in this microcontroller namely PORT A, PORT B, PORT C, PORT D and PORT E. Among these ports PORT B, PORT C and PORT D contains 8-pins, where PORTA contains 6-pins and PORT E contains 3-pins.

Each pin in the ports can be used as either input or output pins. Before using the port pins as input or output, directions should be given in TRIS register. For example setting all the bits in TRIS D register indicates all the pins in PORT D are used input pins. Clearing all the bits in TRIS D register indicates all the pins in PORT D are used as output pins. Likewise TRIS A, TRIS B, TRIS C, TRIS E registers available for PORT A, PORT B, PORT C and PORTE. Other than the normal Microcontrollers PIC family supports more features, so we have PIC 16F877 as the main controller. The Main Features and Peripherals are discussed below.

### c. Memory organization

There are three memory blocks in each of the PIC16F87X MCUs. The Program Memory and Data Memory have separate buses so that concurrent access can occur and is detailed in this section.

### V. USART

Computers must be able to communicate with other computers in modern multi processor distributed systems. One cost-effective way to communicate is to send and receive data bytes serially. The P16F77 micro controller has serial data communication circuits that transmits and receive the data. By configuring the **SPEN** in **RCSTA** we can enable the serial communication in P16F877. Similarly by configuring the **TXSTA** (Transmit Status and Control Register) and **RCSTA** (Receive Status and Control Register) we can transmit and receive the data serially. The any date in **TXREG** (Transmit Register) transmit the data and the data which is in **RCREG** (Receive Register) is the receive data. The following figure shows the TXSTA and RCSTA,

### 1. USART Baud Rate Generator (BRG)

The BRG supports both the Asynchronous and Synchronous modes of the USART. It is a dedicated 8-bit baud rate generator. The SPBRG register controls the period of a free running 8-bit timer. In Asynchronous mode, bit BRGH (TXSTA<2>) also controls the baud rate. In Synchronous mode, bit BRGH is ignored. Table 10-1 shows the formula for computation of the baud rate for different USART modes which only apply in Master mode (internal clock).Given the desired baud rate and FOSC, the nearest integer value for the SPBRG register can be calculated using the formula . From this, the error in It may be advantageous to use the high baud rate (BRGH = 1), even for slower baud clocks. This is because the FOSC/(16(X + 1)) equation can reduce the baud rate error in some cases. Writing a new value to the SPBRG register causes the BRG timer to be reset (or cleared). This ensures the BRG does not wait for a timer overflow before outputting the new baud rate

| SYNC | BRGH = 0 (Low Speed) | BRGH = 1 (High Speed) |
|---|---|---|
| 0 | (Asynchronous) Baud Rate = Fosc/(64(X+1)) | Baud Rate = Fosc/(16(X+1)) |
| 1 | (Synchronous) Baud Rate = Fosc/(4(X+1)) | N/A |

X = value in SPBRG (0 to 255)

Table 1. Band Rate Formula

### 2. USART Asynchronous mode

In this mode, the USART uses standard non-return-torero (NRZ) format (one START bit, eight or nine data bits, and one STOP bit). The most common data formats 8-bits. An on-chip, dedicated, 8-bit baud rate generator can be used to derive standard baud rate frequencies from the oscillator. The USART transmits and receives the LS b first. The transmitter and receiver are functionally independent, but use the same data format and baud rate. The baud rate generator produces clock, either x16 or x64 of the bit shift rate, depending on bit BRGH (TXSTA<2>). Parity is not supported by the hardware, but can be implemented in software (and stored as the ninth data bit). Asynchronous mode is stopped during SLEEP.

Asynchronous mode is selected by clearing bit SYNC (TXSTA<4>).The USART Asynchronous module consists of the following important elements:

- Band Rate Generator
- Sampling Circuit
- Asynchronous Transmitter
- Asynchronous Receiver

### 3. USART Asynchronous transmitter

The heart of the transmitter is the transmitter (serial) shift register (TSR). The shift register obtains its data from the read/write transmit buffer, TXREG. The TXREG register is loaded with data in software. The TSR register is not loaded until the STOP bit has been transmitted from the previous load. As soon as the STOP bit is transmitted, the TSR is loaded with new data from the TXREG register (if available). Once the TXREG register transfers the data to the TSR register occurs in one TCY), the TXREG register is empty and flag bit TXIF (PIR1<4>) is set. This interrupt can be a bled/disabled by setting/clearing enable bit TXIE (PIE1 <4>). Flag bit TXIF will be set, regardless of the state of enable bit TXIE and cannot be cleared in software. It will reset only when new data is loaded into the TXREG register.

While flag bit TXIF indicates the status of the TXREG register, another bit TRMT (TXSTA<1>) shows the status of the TSR register. Status bit TRMT is a read only bit, which is set when the TSR register is empty. No interrupt logic is tied to this bit, so the user has to poll this bit in order to determine if the TSR register is empty .Transmission is enabled by setting enable bit TXEN(TXSTA<5>). The actual transmission will not occur until the TXREG register has been loaded with data and the baud rate generator (BRG) has produced a shift clock (Figure 10-2). The transmission can also be started by first loading the TXREG register and then setting enable bit TXEN. Normally, when transmission is first started, the TSR register is empty. At that point, transfer to the TXREG register will result in an immediate transfer to TSR, resulting in an empty TXREG. Aback-to-back transfer is thus possible Clearing enable bit TXEN during a transmission will cause the transmission to be aborted and will reset the transmitter. As a result, the RC6/TX/CK pin will revert to hi-impedance. In order to select 9-bit transmission, transmit bit TX9 (TXSTA<6>) should be set and the ninth bit should be written to TX9D (TXSTA<0>). The ninth bit must be written before writing the 8-bit data to the TXREG register .This is because a data write to the TXREG register can result in an immediate transfer of the data to the TSR register (if the TSR is empty). In such a case, an incorrect ninth data bit may be loaded in the TSR register

### 4. USART Asynchronous receiver

The data is received on the RC7/RX/DT pin and drives the data recovery block. The data recovery block is actually a high speed shifter, operating at x16 times the baud rate; whereas, the main receive serial shifter operates at the bit rate or at FOSC. Once Asynchronous mode is selected, reception is enabled by setting bit CREN (RCSTA<4>).The heart of the receiver is the receiver (serial) shift register data in the RSR is transferred to the RCREG register (if it is empty). If the transfer is complete, flag bit RCIF (PIR1<5>) is set. The actual interrupt can be enabled/disabled by setting/clearing enable bit RCIE (PIE1<5>). Flag bit RCIF is a read only bit, which is cleared by the hardware. It is cleared when the RCREG register has been read and is empty. The RCREG is a double buffered register (i.e., it is a two deep

FIFO). It is possible for two bytes of data to be received and transferred to the RCREG FIFO and a third byte to begin shifting to the RSR register. On the detection of the STOP bit of the third byte, if the RCREG register is still full, the overrun error bit OERR (RCSTA<1>) will beset. The word in the RSR will be lost. The RCREG register can be read twice to retrieve the two bytes in the FIFO. Overrun bit OERR has to be cleared in software. This is done by resetting the receive logic (CREN is cleared and then set). If bit OERR is set, transfers from the RSR register to the RCREG register are inhibited, and no further data will be received. It is therefore, essential to clear error bit OERR if it is set. Framing error bit FERR (RCSTA<2>) is set if a STOP bit is detected as clear. Bit FERR and the 9th receive bit are buffered the same way as the receive data.

### 5. Reset

The PIC16F877 has various kinds of reset.

1) Power-on Reset(POR)
2) MCLR Reset during normal operation
3) MCLR Reset during SLEEP
4) WDT Reset (during normal operation)
5) WDT Wake-up (during SLEEP)
6) Brown-out Reset (BOT)

Some register are not affected in any RESET condition. Their status is unknown on POR and unchanged in any other RESET. Most other registers are reset to a "RESET" state on Power on Reset (POR), during SLEEP.

### VI. Future Scope

1) It can be extended and used for three phase motors.
2) The size of paper can be reduced by using appropriate electronic components.

### A. Advantages

- It involves transmission of data through inductive flux without any contact between two modules.
- Single system can be used with multiple clients.
- This reduces a lot of hazards such as interference with other modules and RF transmitters.
- It is also used to find faults in all types of single phase AC motors.

### B. Future improvements : Improve Power Transfer Range

Mounting the coils closer to the hinges could increase the range of the existing power transfer system. Mounting the coils in this area was avoided in this paper because of the shielding effect of the metal shell of the door. With modification to the refrigerator structure however, a portion of the metal sheet could be removed to accommodate the power transfer coils. With careful coil positioning, power transfer may be achieved for around 60% of the range of the door. Mounting the coils closer would improve the magnetic coupling and therefore increase the power transfer. It was found that the air cores had a good tolerance to varying the angles between them; therefore this may be a better solution if modification to the refrigerator structure is allowed for in the design.

### C. Increase Data Rates

Although high data rates were not required for this paper, it may be well worth investigating the limitations of the designed IPT communication system. The communication system is a relatively cheap solution to contact less two-way communication across air gaps of up to 10cm. If high data rates are achieved, this type of communication system may find uses in future applications. At present the main constraint on the data rate was the low pass filter used - a simple resistor and capacitor combination - to keep the costs low and an adequate solution for this application. However, if a 2 or 3 order filter was designed and implemented, an immediate improvement in data rate capability would be observed. Another possible modification to increase the data rates would be to increase the transmission frequency. This would increase the number of cycles of the sinusoidal transmission waveform per data bit sent and hence enable the filter to be more effective.

### VII. Conclusion

This approach is used for indicating various parameters of single phase AC motors using the help of IPT and PIC Microcontroller (16F877). It is also used to indicate the faults. The output is displayed by using LCD. Unlike other wireless techniques such as RF and IR this method demonstrates a simple non tunable means of communication that is interference free and can be used with power transfer techniques. A contact less, two-way communication system was designed and implemented across an air gap, achieving data rates of 250 bits/second.

All the components in this paper works satisfactory. It is one of the cheapest methods which are suitable for all types of single phase motors. A self-switching power supply and self-regulating pickup were implemented, resulting in a completely independent power supply system requiring no input from the microcontroller. All of the specifications for the paper were met resulting in a reliable and robust solution to the problem. A new, innovative concept in IPT data transmission was introduced and proven to be successful.

Author's Profile

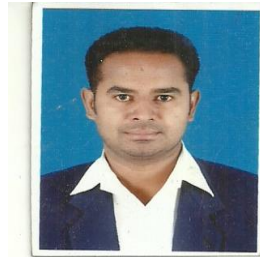

**K.P.Shaji**, received B.E.S from Sengunthar Arts & Science College, University of Madras, 1995. M.Sc Applied Electronics from SNR & Sons college Coimbatore, Bharathiyar University, 1997. M.Phil in Bharathidasan University, 2006. Currently he is working as HOD cum Associate Professor in the Department of Electronics & Communications, Muthayammal College of Arts & Science, Namakkal. His area of interests are Digital Electronics & Analaog Digital Communications.

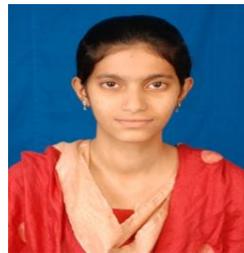

**I.Alsheba**, Pursuing her Bachelor's in Electronics & Electricals Engineering in KSR College of Technology, Anna University, Namakkal, India. Her area of interest are Power Electronics.

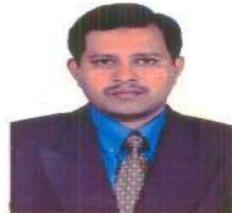

**Y.A. Syed Khadar**, received B.E.S from Sengunthar Arts & Science College, University of Madras, 1996. M.Sc Applied Electronics from Bharathiyar University, 1998. M.Phil in Bharathidasan University, 2008 & Currently pursuing his Ph.D Electronics in Bharathiyar University from 2012. Currently he is working as an Associate Professor in the Department of Electronics & Communications, Muthayammal College of Arts & Science, Namakkal. His area of interest are Micro Processor & Controller.

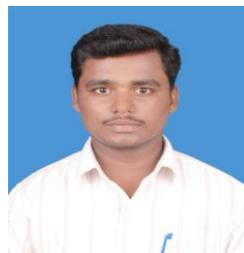

**S.Kannan**, received B.SC Electronics & Communication (2006), M.Sc Electronics & Communication (2008) from Muthayammal College of Arts & Science, Periyar University, Namakkal. Currently he is working as an Asst.Professor in the Department of Electronics & Communications, Muthayammal College of Arts & Science, Namakkal. His area of interests are Digital Electronics & Analaog Digital Communications.